# Square ice in graphene nanocapillaries

G. Algara-Siller[1], O. Lehtinen[1], F.C. Wang[2], R. R. Nair[3], U. Kaiser[1], H. A. Wu[2], I. V. Grigorieva[3], A. K. Geim[3]

[1]Central Facility for Electron Microscopy, Group of Electron Microscopy of Materials Science, University of Ulm, 89081 Ulm, Germany

[2]Chinese Academy of Sciences Key Laboratory of Mechanical Behavior & Design of Materials, Department of Modern Mechanics, University of Science & Technology of China, Hefei, Anhui 230027, China

[3]School of Physics & Astronomy, University of Manchester, Manchester, M13 9PL, UK

**Adsorbed layers of water are ubiquitously present at surfaces and fill in microscopic pores, playing a central role in many phenomena in such diverse fields as materials science, geology, biology, tribology, nanotechnology, etc. Despite such importance, the crystal structure of nanoconfined water remains largely unknown. Here we report high-resolution electron microscopy of mono- and few- layers of water confined between two graphene sheets, an archetypal example of hydrophobic confinement. Confined water is found to form square ice at room temperature – a phase with symmetry principally different from the conventional tetrahedral geometry of hydrogen bonding. The square ice has a high packing density with a lattice constant of $\approx$2.83 Å and during TEM observation assembles in bi- and tri- layer crystallites exhibiting AA stacking. Our findings are important for understanding of interfacial phenomena and, in particular, shed light on ultrafast transport of water through hydrophobic nanocapillaries. Our MD simulations suggest that square ice is likely to be common inside hydrophobic nanochannels, independent of their exact atomic makeup.**

Three-dimensional (3D) water exists in many forms, as liquid, vapor and as many as 15 crystalline and some amorphous phases of ice, with the commonly found hexagonal ice alone being responsible for the fascinating variety of snowflakes [1,2]. Less noticeable but equally ubiquitous is water present at interfaces and in microscopic pores where nanometer-scale confinement makes the structure of water and its dynamics radically different from bulk water [3,4]. Confined and interfacial water has attracted dedicated interest in fields ranging from life to earth to materials sciences, playing a crucial role in such diverse phenomena as protein assembly, nanofriction, filtration, dissolving, frost heaving, detergent cleaning, heterogeneous catalysis and so on [5-8]. It is now well established that near a solid surface, whether hydrophilic or hydrophobic, water forms a layered structure made up of distinct monolayers [3-5,9-22]. However, the structure within these layers remains largely unknown. Molecular dynamics (MD) simulations [9-17] predicted a great variety of phases, although the results are sensitive to modelled conditions and some seem conflicting. For example, a buckled monolayer ice was found inside hydrophilic nanochannels [11] and a flat hexagonal ice inside hydrophobic ones below room temperature [12,15,22]. On the other hand, no in-plane order was observed inside mica (hydrophilic) and graphite (hydrophobic) nanochannels at and above room temperature [11,14]. A close analogue of planar square ice was reported in MD simulations of water inside carbon nanotubes [9,10,17]. In this case, water molecules form a monolayer that can be viewed as a sheet of square ice rolled-up into a quasi-1D cylinder. Experimentally, neutron studies [17] have shown features consistent with the existence of such 'ice nanotubes' filled with a 1D water chain, which melt above 50 K. 2D ices have also been found and investigated at the surfaces of mica and graphite [18-22]. The studies using scanning probe microscopy [19-21] and electron crystallography [18,22] showed that near-surface water forms correlated, solid-like layers. As for their internal structure, information is only available for 2D ices grown below 150 K, which are found to be hexagonal, with in-plane coordination similar to bulk ices [18,22].

We report the microscopic structure of mono- and few- layers of water confined between graphene sheets using aberration-corrected high-resolution transmission electron microscopy (TEM). The TEM samples were prepared using procedures described in Supplementary Information [23]. In brief, a graphene monolayer was deposited onto the standard TEM grid, exposed to water and covered with another graphene crystal [24,25]. Most of the water was squeezed out by van der Waals (vdW) forces that brought the two graphene sheets together but some adsorbed water remained trapped in numerous pockets of a submicron size (Fig. S1). For comparison, we also prepared reference samples, similar to those described in ref. 24, in which a larger amount of deposited water resulted in three dimensional (3D) droplets of ~10 to 100 nm in thickness (Fig. S2). The prepared samples were studied in a transmission electron microscope (FEI TITAN 80-300) with an imaging side aberration corrector, operated at 80 kV.

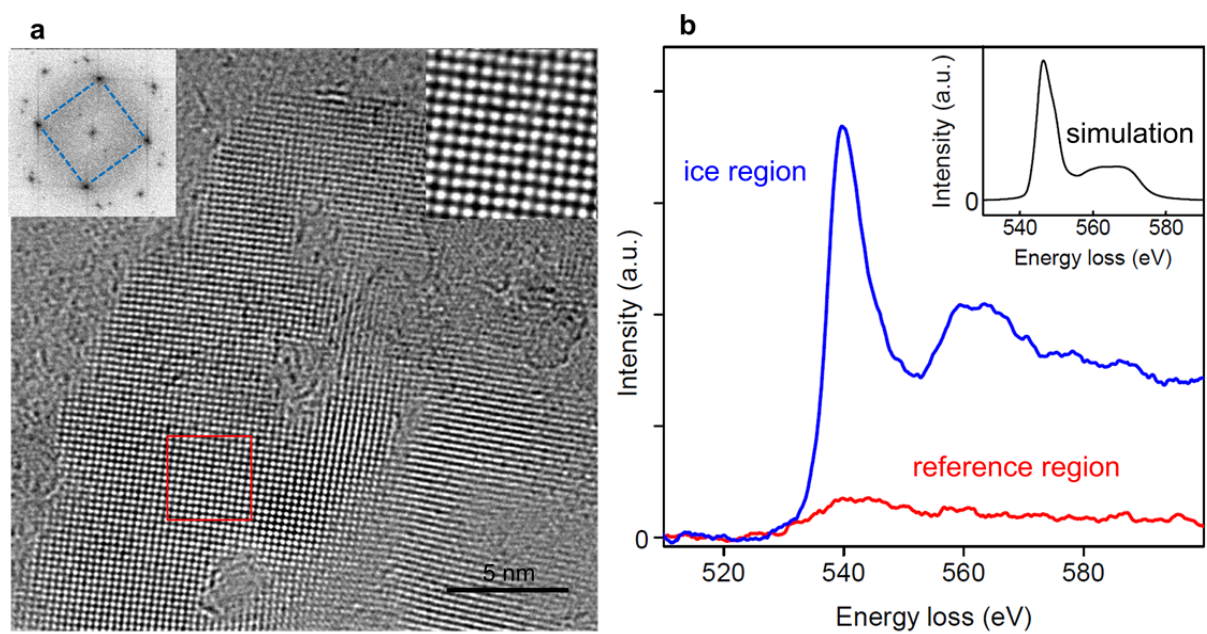

*Figure 1. Square ice. (a)* *A part of a large water pocket (see Fig. S1a). Several such pockets were studied. Top right inset: Magnified image of the ice lattice in the area outlined in red in the main image. Top-left: Fourier transform of the entire image shows four first-order maxima of the square lattice; the square symmetry is highlighted by the dashed blue lines; the two hexagonal sets come from the encapsulating graphene layers and show no alignment with the ice lattice. Irregular structures in the main image are hydrocarbon contamination.* *(b)* *Oxygen K-edge EELS obtained from areas containing the 2D ice and no visible water (blue and red curves, respectively). Both spectra come from similar size areas (~100 nm in diameter). Inset: Simulated EELS for $I_c$ ($I_h$ exhibits a very similar spectrum), see ref. 28.*

Typical atomic-resolution images of graphene-confined water are shown in Fig. 1a and Fig. S1b. High-contrast dark spots correspond to oxygen atoms that indicate positions of water molecules. Hydrogen atoms give too little contrast to be resolved even by the state-of-the-art TEM. The encapsulating graphene layers are seen in some parts of the images as faint background with hexagonal symmetry [25,26]. Further improvement of the TEM images can easily be achieved by digitally subtracting the contribution from encapsulating graphene [25] but, given the high quality and contrast of the raw data, we avoided this additional processing. One can see that water molecules form a regular square lattice. Its analysis by the fast Fourier transform yields the distance $a$ between the nearest oxygen atoms of 2.83±0.03Å (inset of Fig. 1a). No alignment between the water and graphene lattices could be found in the images. It is instructive to mention that the atomic-resolution imaging of interfacial ice has become possible uniquely due to the use of graphene: its low atomic number and crystallinity allow a minimal background and high contrast for oxygen atoms [24,26] whereas graphene's mechanical strength, high thermal and electrical conductivity and chemical stability ensure protection of encapsulated water from sublimation and effects of electron irradiation [23,25,27].

To confirm that the observed square lattice is indeed formed by water molecules, we acquired electron energy loss spectra (EELS). Typical spectra around the oxygen K-edge, which are sensitive to states of water [28], are shown in Fig. 1b. The figure compares EELS from areas such as in Fig. 1a and areas without any visible amount of trapped water. Little oxygen signal comes from the reference region whereas the ice region exhibits the spectrum that has an overall shape qualitatively similar to EELS for 3D ices such as hexagonal $I_h$ and diamond cubic $I_c$ ices [28]. However, there are differences, too (inset of Fig. 1b). The main EELS peak is shifted by ≈6 eV, and a secondary peak occurs around 560 eV. The separation between the main and secondary peaks allows an estimate of the oxygen separation, which yields ≈2.8 Å (see section 'EELS analysis' in [23]), in agreement with the observed lattice constant. In addition, we measured EELS for large water droplets found in our reference samples. The latter exhibited spectra typical for liquid water-vapour mixtures (Fig. S2b).

The square ice is found to be highly mobile under the electron beam as illustrated in Fig. 2 and Fig. S3. After the first few seconds, a uniform monolayer starts breaking up into ~10 nm crystallites with sharp crystallographic edges. They change configuration, merge and break up again with bi- and tri- layer crystallites becoming commonly present. The frequency of jumps and reconstruction increases with the beam current. As the crystallites move and coalesce, grain boundaries and dislocations are formed but the high crystallinity is preserved with no sign of ice melting or amorphization. Strong variations in TEM contrast seen, for example, in Fig. 2 correspond to different numbers of layers. The layered structure of the square ice is elucidated in Fig. 3, where the three equal-height steps in contrast correspond to the thickness changing from mono- to bi- and to tri- layer ice (see Fig. S4 for details of this analysis). By using the quantified oxygen contrast to examine many obtained images and videos of the 2D ice, we have found that three was the maximum number of layers observed in our experiments.

TEM images such as that in Fig. 3a allow unambiguous determination of the stacking order in the few-layer ice. We find that its water molecules are AA stacked, that is, oxygen atoms in adjacent layers are sitting directly on top of each other. The AA stacking is clear from comparison of Fig. 3a with simulated images in Figs. 3b,c. In the case of AB stacking, oxygen atoms in the second layer would occupy the sites in between the oxygen sites in the first and third layers, resulting in a perceived 45° rotation of the water lattice and a reduced projected oxygen spacing (≈2.0 Å). Furthermore, the reduced spacing would lead to a lower contrast and, consequently, a qualitatively different appearance of AB bilayers (Fig. 3c) compared to the experimental images (Fig. 3a). Note that, in bulk water, the stacking order controls the difference between different ices. For example, hexagonal and diamond cubic ices are both formed by puckered hexagonal layers but exhibit AB and ABC stacking, respectively. In both cases, bonding between water molecules follows so-called 'ice rules' that require a tetrahedral coordination of hydrogen bonds [1,2]. In contrast, the observed few-layer ice corresponds to 90° hydrogen bonding within and between layers.

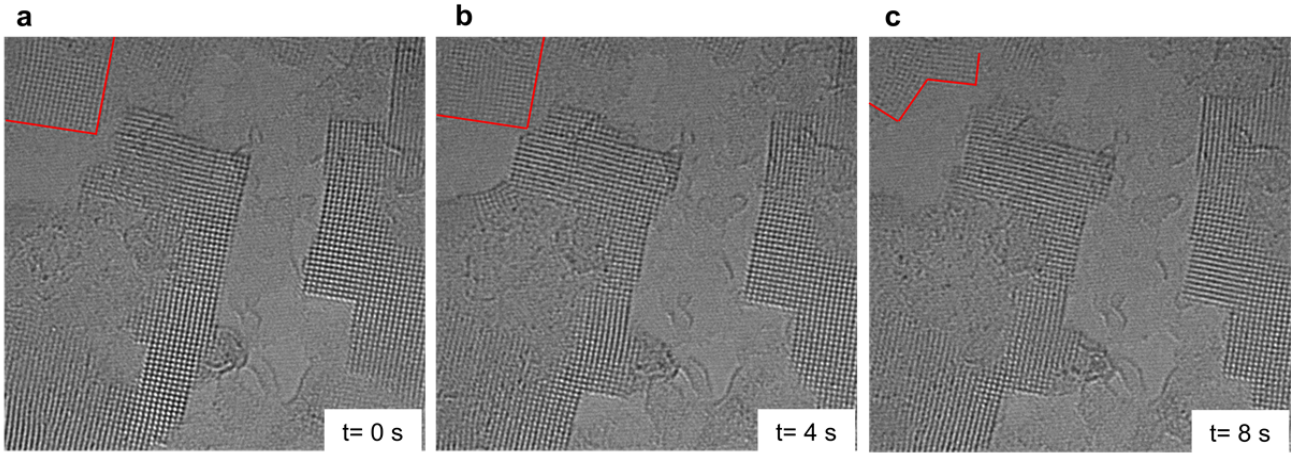

*Figure 2. Dynamics of 2D ice crystallites. Numerous changes occurred during the time elapsed between the snapshots extracted from a TEM movie (Supporting video). Red lines highlight some of the changes: bilayer crystallite in (a) thins down to a monolayer in (b), then splits into two crystals separated by a grain boundary in (c). More snapshots from the same sequence are shown in Fig. S3.*

To support our TEM observations, we have carried out MD simulations of water confined in graphene nanocapillaries [23]. The distance between the centres of carbon planes was varied from 6.5 to 11.5 Å to allow one to three monolayers of water inside (Fig. S5). For narrow 2D channels that can accommodate only one monolayer, we always find square ice with $a$ =2.81±0.02 Å, in excellent agreement with the experiment (Fig. 3d and Fig. S6). This MD result is robust, showing little dependence on the capillary width, applied pressure and whether graphene sheets were made rigid, flexible or freely moving [23].

However, for wider capillaries that allow two or three monolayers of water, no in-plane ordering could be found under simulated ambient conditions. Instead, water molecules make a distinct layered structure but maintain locally the tetrahedral arrangement of hydrogen bonds both within and between the layers (Figs. S7a,b). To understand this apparent disagreement with the observed few-layer ice, we believe it is important to take into account a pressure that adhesion between encapsulating graphene sheets imposes on water by trying to squeeze it into a smaller volume. As illustrated in Fig. S8, this pressure - we refer to it as van der Waals - can be estimated from the energy gain due to such squeezing, which yields $P_W \approx E_W/d$ ~1 GPa, where $E_W \approx$ 20-30 meV/Å$^2$ is the adhesion energy [29,30] and $d \approx$ 3.5 Å, a typical interlayer distance. This estimate agrees with our MD simulations of the hydrostatic pressure acting on He gas trapped between freely moving graphene sheets (section 'Van der Waals pressure' in [23]). To mimic the effect of vdW pressure that is integral for our nanocapillaries, we have applied an external hydrostatic pressure $P$ as illustrated in Fig. S5. A pronounced transition is found from in-plane disorder at low $P$ to layers of square ice at $P$ >1 GPa (Figs. S7,S9). The layered ice exhibits the same lattice constant ($a$ =2.81±0.02 Å) as the monolayer, in agreement with the experiment.

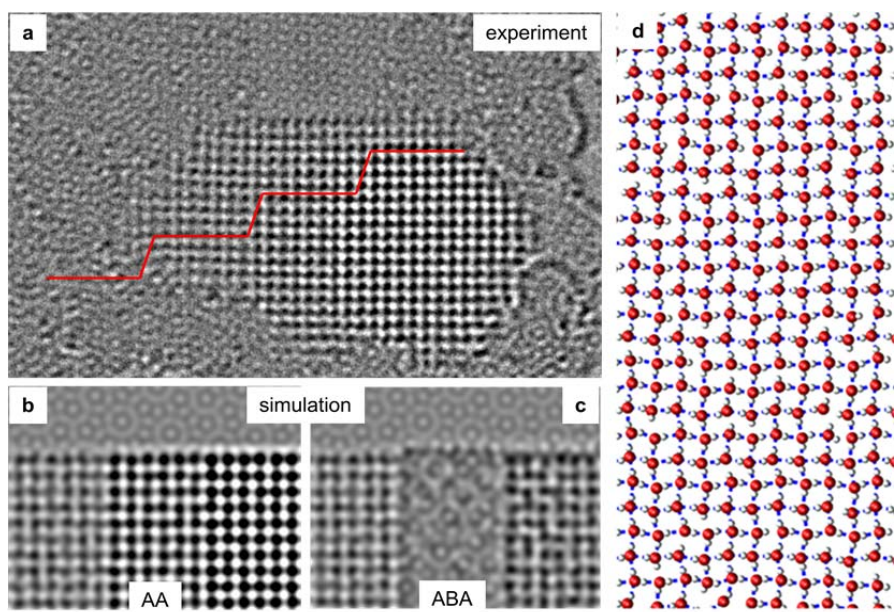

***Figure 3. Few-layer ice and its stacking order. (a)*** *Isolated ice crystal with a varying number of layers. The moiré pattern (seen most clearly at the top) is due to encapsulating graphene sheets. The red line shows the contrast averaged over the corresponding parts of the image (Fig. S4). The contrast changes in quantized steps, which allows unambiguous identification of the number of layers in ice crystals.* ***(b-c)*** *Simulated TEM images of mono-, bi- and tri- layer ice with AA and AB stacking, respectively. The simulated image of AA stacked ice (panel b) yields a linear dependence of contrast on the number of layers, in agreement with the equal-height steps in panel (a).* ***(d)*** *Typical MD snapshot of water in a graphene nanocapillary.*

On the other hand, our MD simulations have failed to reproduce the observed AA stacking. They show a tendency to form crystals with AB staking or no interlayer order (Figs. S7,S9). The disagreement remains to be understood but perhaps is not surprising. As $P$ increases to reach the crystallization transition at ~1 GPa, hydrogen bonds switch to the in-plane configuration (Figs. S7d, S9b) so that the coupling between monolayers of square ice becomes vdW-like. Such weak interlayer coupling is known to be notoriously difficult to accurately account for in theoretical analysis. Moreover, in the experiment the graphene confinement is terraced with sharp steps between ice terraces (see Fig. 3), which can result in extra lateral forces acting on different layers.

Finally, our simulations for bi-/tri-layer ice yield an interlayer separation $c$ =2.8±0.3Å, practically independent of pressure and other parameters. This means that the few-layer ice attains a crystal lattice close to the simple cubic structure ($c = a$) and its density is ≈1.5 times higher than that of the common ice, $I_h$. The exact structure is still likely to be - strictly speaking - tetragonal because of the qualitative difference between inter- and intra- layer bonding (Figs. S7,S9).

To conclude, the reported room-temperature ice exhibits square planar coordination, qualitatively different from the conventional tetrahedral coordination of water molecules observed in bulk and surface ices. We expect the square ice to be common under ambient conditions inside hydrophobic nanocapillaries, basically because water-surface interaction for such confinement is by definition much weaker than the interaction between water molecules. Indeed, our MD simulations yield the same square lattices inside non-graphene capillaries with widely-varying hydrophobicity (Fig. S10). The existence of low-dimensional ice at room temperature has been central for explaining ultrafast permeation of water through hydrophobic nanocapillaries including carbon nanotubes, protein nanopores and graphene-based membranes. This explanation involves correlated, train-like movement of ordered water, and our report supports this idea. Our results also indicate that the invoked van der Waals pressure can generally be important at the nanoscale.

# Supplementary Information

#1 Sample preparation

Graphene monolayers were grown on Cu foils by chemical vapour deposition and then transferred onto gold Quantifoil grids (Au mesh covered with an amorphous carbon film having a dense array of holes with a diameter of 1.2 μm). Importantly, a thin layer of Pt was deposited on the TEM grids prior to graphene transfer in order to reduce hydrocarbon contamination [31]. For the transfer, several Quantifoil grids were immersed in isopropanol and then placed 'face down' onto a ~1 $cm^2$ piece of graphene on copper. As isopropanol evaporated, the amorphous carbon film of the TEM grid came in contact with graphene and became attached to it [32]. After that the assembly was floated on the surface of ammonium peroxodisulfate for several hours until all copper slowly etched away, leaving free-floating graphene with several TEM grids attached to it. This was broken into pieces, with just one grid attached to each piece, washed in water and isopropanol and left to dry in air at room temperature. After that 1 μL of deionized water was carefully cast on top of one of the grids, and finally the second graphene-covered grid was placed on top, covering the water droplet. The resulting sandwich was left to dry under ambient conditions overnight, during which time the water drop slowly evaporated, bringing the two graphene layers together and gradually squeezing the liquid out so that only a small amount of adsorbed water remained captured in between the graphene sheets.

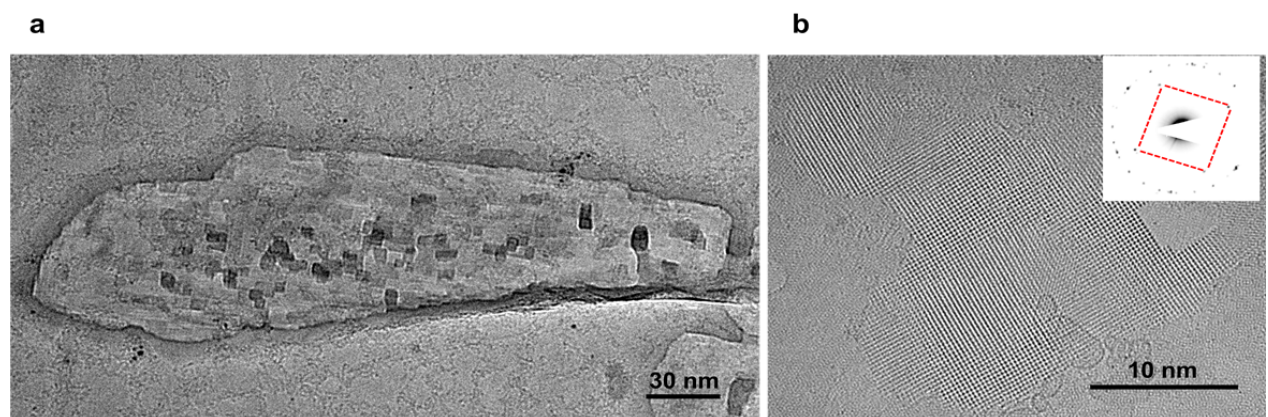

*Fig. S1. 2D water captured between graphene monolayers. (a) Confined water imaged at low magnification. Typical lateral size of such water pockets was ~100 nm. The square pattern reveals a collection of crystallites formed by water molecules. (b) Another example of high-magnification image of 2D ice. The inset shows selected area electron diffraction (SAED) from a confined water region of ~120 nm in diameter. The four diffraction spots from the square ice lattice (connected by the red lines for clarity) yield the lattice parameter of 2.83±0.03Å, same as the value obtained using fast Fourier transform in Fig. 1a of the main text.*

#2. Transmission electron microscopy

The use of both a relatively low operating voltage (80 keV) and the aberration correction were essential in our study: higher voltages are known to produce knock-on damage in graphene [33] whereas the aberration correction even at relatively low acceleration voltage allows atomic resolution imaging of both graphene and water. The spherical aberration coefficient was set to 30 μm and the sample was imaged at Scherzer focus, resulting in dark atom contrast. The vacuum level in the microscope was <$10^{-8}$ mbar and all experiments were done at room temperature. Aberration-corrected high-resolution TEM images were acquired at $3\times10^4$ e$^-$/nm$^2$ dose and exposure time of ~1s per frame. The exposure of the studied area prior to acquisition of the reported images was approximately 600 e$^-$/nm$^2$. No notable heating by the electron beam is expected under these operating conditions: For amorphous carbon films under experimental conditions similar to ours, the heating was estimated to be ~1 K [ref. 34]. As thermal conductivity of graphene is 3,000 times higher than that of amorphous carbon, we expect even less heating in our case. This is in agreement with our observation that the appearance of the square ice remained qualitatively the same under increasing/decreasing the beam current.

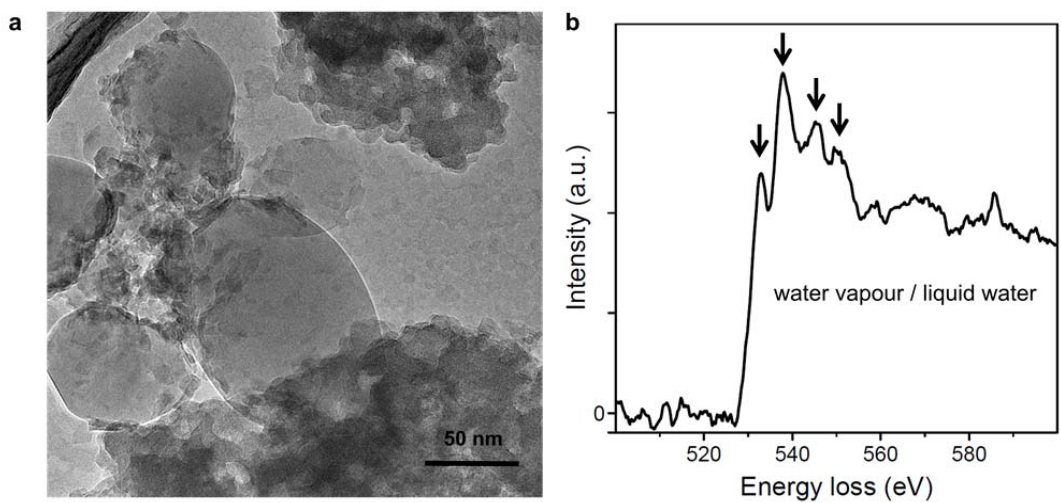

*Fig. S2. Large water droplets and their EELS. (a)* Low-magnification TEM image of a reference sample where water formed large bubbles (diameter of ~100 nm). *(b)* Spectrum near the O K-edge from the droplets shown in (a). The four marked peaks at 532, 537, 545 and 550 eV are in agreement with the previously studied spectra for mixtures of water vapour and liquid water [39,40].

The majority of our samples remained intact during the relatively long exposures to the electron beam (typically, 10-20 min), and only in some cases etching of graphene was observed [27,33-36]. At first sight, it may be surprising that the reported ice crystals survive such relatively long imaging, in contrast to earlier work on frozen aqueous samples that always suffered from fast sublimation under the electron beam [27,28,37]. The high stability of 2D ice in our work is due to graphene encapsulation: The highly conductive graphene layers (both electrically and thermally) provide an efficient channel for absorbing and rapidly dissipating the energy introduced by the electron impacts, thus reducing the damage to ice crystals. This effect of graphene encapsulation was recently demonstrated in studies [25,38] of the electron beam damage of monolayer MoS$_2$, a radiation sensitive material. The encapsulation reduced the damage rate by nearly three orders of magnitude as compared to non-encapsulated MoS$_2$. Accordingly, graphene capillary in the present work served not only as a confinement channel but also as protection against the beam damage, which enabled us to observe ice crystals without immediately evaporating/destroying them.

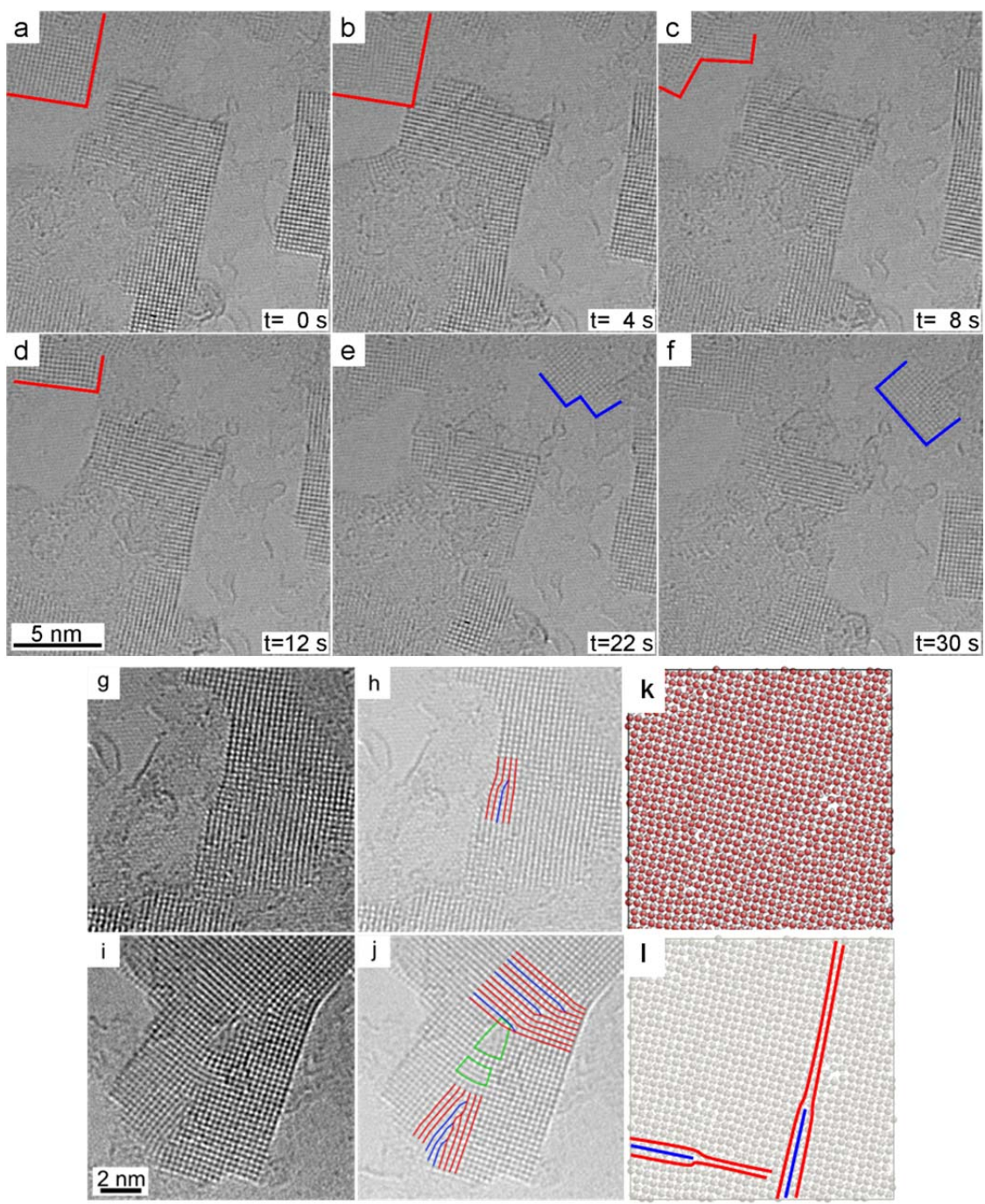

***Fig. S3. Dynamics of 2D ice crystallites and formation of crystal defects.*** *(**a-f**) High resolution TEM snapshots of the same area observed over 30 s, demonstrating continuous reorganization of ice crystallites. The top row (a-c) is explained in Fig. 2 of the main text. In (d), a trilayer crystal is formed in the same area marked in red. In (e), a new crystallite, outlined in blue, appears in the top right corner, growing and propagating towards the center in (f). Also, the overall shape of the larger crystallites is changing from panel to panel. (**g-j**) Examples of an edge dislocation (g) and tilt grain boundaries (i) in 2D ice. (h) and (j) are the same images as (g) and (i), respectively, but with reduced contrast; atomic rows overlaid with red and blue lines to highlight the defects. Red lines mark existing atomic rows; blue lines mark extra rows originating from dislocations; green shapes outline defect structures, where exact row arrangements are indiscernible. (**k,l**) Monolayer ice found in MD simulations also shows dislocations, indicating that they are intrinsic to the formation of 2D ice at room temperature.*

The high mobility of the 2D ice under the electron beam is demonstrated by the snapshots in Fig. S3 (Fig. 2 in the main text shows a part of the same sequence). As the crystallites move and coalesce into larger crystals during TEM observation, grain boundaries and dislocations are formed, which is typical for polycrystals.

#3. EELS analysis.

Electron energy loss spectra [39,40] were acquired using GIF Quantum ER filter in the diffraction mode (convergence angle of 1.8 mrad and collection angle of 4.4 mrad). The spectra were recorded with exposure and integration times of 5 and 200 s, respectively. A low pass filter (2 px) and linear background subtraction were applied. All EELS showed prominent signals from carbon and oxygen. Hydrogen is not detectable with the instrument. No other elements could be detected, except for a small Si peak that is commonly present in all areas because of contamination during sample preparation [41].

The overall shape of EELS curves allows us to distinguish immediately between solid and liquid/vapour water (cf. Fig. 1 and Fig. S2). Furthermore, we note that bulk ices such as $I_c$ and $I_h$ are expected to have rather similar spectra [28], more similar to each other than to the observed EELS for square ice. This is understandable as both $I_c$ and $I_h$ consist of puckered hexagonal layers, which is different from the planar square configuration. To gain further information from the observed EELS, we have used the spectral positions of the main and secondary peaks (see the main text) to estimate interatomic distances in the square ice. Secondary peaks after the main ionization threshold at ≈540 eV (Fig. 1) arise due to multiple scattering, and the energy difference, $\Delta E$, between them obeys the relation [42] $\Delta E \times R^2 = C$, where $R$ is the separation between oxygen atoms and $C$ is a constant. Using $C$ =150 as empirically determined for cubic ice [43] ($I_c$ is the closest approximation to our 2D ice that can be found in literature) and the energy of ≈559 eV for the prominent secondary peak in the observed EELS of Fig. 1, we find $R$ = 2.80±0.07 Å. The good agreement between this value and the lattice constant $a$ obtained from atomic-resolution TEM images provides an independent confirmation of the crystal structure of the square ice.

In principle, the interlayer spacing in crystals can be determined by their TEM imaging at different tilt angles. Unfortunately, this is impossible to achieve in practice for nanoscale crystals that constantly move and rotate under the electron beam. Indeed, for acquiring interpretable high-resolution images, crystals need to be oriented precisely along a zone axis and remain stationary during imaging. The reason that high-resolution imaging is possible in [001] direction in our case is that the graphene layers fix the orientation of ice crystals relative to the electron beam. Even though the crystals change their in-plane orientation (rotate), the [001] direction remains parallel to the optical axis of the microscope.

#4. Analysis of TEM images

To determine the number of layers in 2D ice crystallites, we quantified the contrast from oxygen atoms using a variance filter, as illustrated step-by-step in Fig. S4. This resulted in contrast maps such as that in Fig. S4d. They show that the average contrast changes in steps of equal height. There are three distinct parts in the ice crystallite in Fig. S4a, comprising one, two and three layers of water molecules. The corresponding average contrast for this crystal is plotted in Fig. 3 in the main text.

The simulated images presented in Fig. 3b,c of the main text were obtained as follows. First, we constructed a square lattice of water molecules arranged in one, two or three layers with different stacking orders. After that, TEM images for different arrangements were simulated using QSTEM software [44] and the parameters corresponding to our experimental conditions (accelerating voltage and the spherical aberration coefficient stated above).

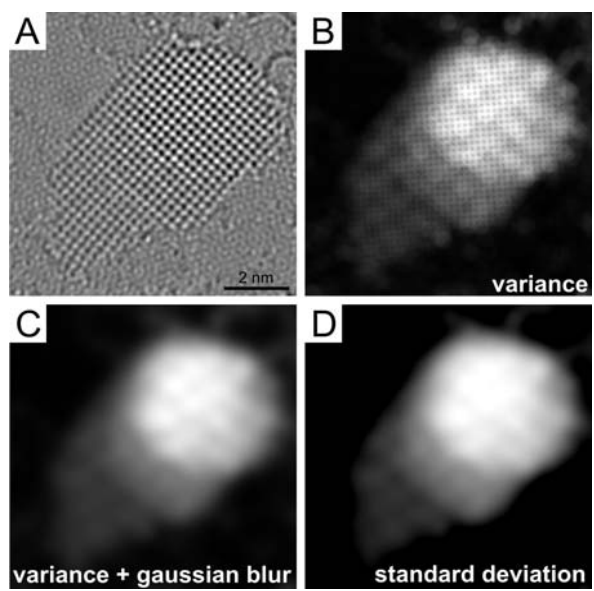

*Fig. S4. Quantifying the contrast to determine the number of layers in ice crystallites.* (**A**) *The original TEM image (also shown in Fig. 3A of the main text).* (**B**) *The filter calculates the local variance of the image which is displayed as a greyscale value.* (**C**) *Due to the varying overlap of the filter mask and the crystal lattice, a lattice pattern remains in the variance image, which is removed by applying Gaussian blur.* (**D**) *The final root-mean-square contrast map (i.e., the standard deviation map) is finally obtained by calculating the square root of each pixel value of the variance image.*

#5. Molecular dynamics setup

Most of our simulations were done for the geometry shown in Fig. S5. This involved two water reservoirs that contained 2,000 molecules each and were connected by a relatively long capillary formed by two parallel graphene sheets. The length and width of the graphene channel were kept at 68 Å and 56 Å, respectively. The height $h$ was chosen as 6.5, 9.0 and 11.5 Å in order to accommodate one, two and three layers of water molecules, respectively. The graphene sheets were either kept rigid or allowed to be flexible during simulations. Unless stated otherwise, water was modelled using the extended simple point charge (SPC/E) model, which is described by the sum of the long-range Coulomb potential and the short range Lennard-Jones potential between the interaction sites [45]. Parameters for water/graphene interactions were taken as in ref. 46. The long-range electrostatic interactions were computed using the particle-particle particle-mesh (PPPM) algorithm, with a convergence parameter of $10^{-4}$. Periodic boundary conditions were applied in all three directions of the simulation box.

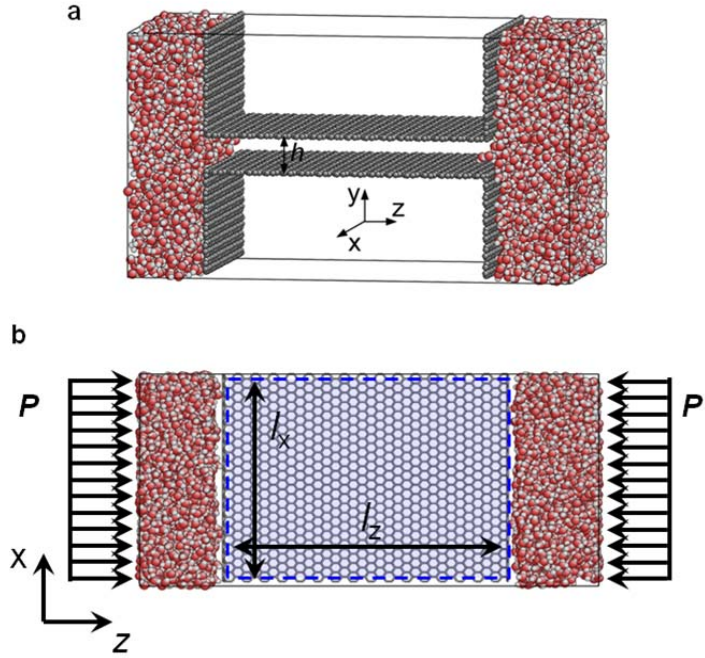

*Fig. S5. MD simulations setup. (a) Initial configuration. (b) Top view of the empty graphene channel, also showing how the external pressure P, needed to mimic the vdW pressure, was applied.*

For monolayer ice, we have also performed MD simulations for freely moving graphene sheets that 'self-consistently' enclose trapped water molecules (Fig. S8d). For bi- and tri-layer ice, the graphene confinement observed experimentally is terraced with atomically-sharp steps between ice terraces (see Fig. 3). It has proven difficult to reproduce such a confinement in MD analysis. Therefore, the high pressure induced by encapsulating graphene sheets (see further) was modelled to a first approximation by applying a hydrostatic pressure $P$ in the direction parallel to the graphene layers as shown in Fig. S5.

The MD simulations were performed in an isothermal-isobaric ensemble, in which the temperature (298 K) and pressure were controlled by the Nose-Hoover thermostat and barostat, respectively. In the equilibrium run, pressure $P$ was kept at 1 atm ($10^{-4}$ GPa) for 5 ns, during which time the water molecules filled the graphene nanocapillary. After that, $P$ was increased to values up to 10 GPa during 15 ns. A time step of 1.0 fs was used for the velocity-Verlet integrator. All the simulations were carried out using LAMMPS [47]. To determine the lattice parameter, we counted the number of water molecules over the entire area of the graphene channel.

#6. Van der Waals pressure
Attractive vdW forces between two graphene sheets favour their adhesion over a maximal area. If a material (for example, a gas bubble) is trapped between the sheets, the bubble will continue to shrink in size until a built-up hydrostatic pressure is able to balance the adhesion forces. For a quasi-2D confinement such as shown in Fig. S8a, it is straightforward to estimate the resulting vdW pressure, $P_W$. Indeed, a displacement $\delta$ of the enclosure boundary results in a gain in adhesion energy equal to $\delta \times L \times E_W$ where $L$ is the enclosure circumference and $E_W$ is the difference between graphene-graphene and graphene-water adhesion energies per unit area. $E_W$ can be estimated as a typical value of the adhesion energy for vdW materials [29,30] because the water-graphene interaction is hydrophobic and relatively small. The above displacement requires the work $F\times\delta$ against the internal pressure, where $F = P_W\times d\times L$. For a monolayer of ice, we can take $d \approx 3.5$ Å, a typical interlayer distance in vdW materials. The equilibrium requires $P_W = E_W/d$, which yields ~1 GPa for $E_W \approx 30$ meV/Å$^2$ found experimentally for monolayer graphene [30].

To support this estimate, we have also 'measured' the vdW pressure in MD simulations. To this end, a small amount of helium gas (1,000 atoms) was confined between freely-moving graphene sheets and the system was allowed to reach equilibrium (Fig. S8b,c). The internal pressure was calculated using the ideal gas law $P = Nk_\mathrm{B}T/V$, where $N$ is the number of He atoms, $V$ the equilibrium volume and $k_\mathrm{B}T$ the thermal energy. The MD results yield a hydrostatic He pressure of $\approx$1.0 GPa, in good agreement with the above estimate. This analysis suggests that high internal pressures are intrinsic for flexible nanoscale confinement and should generally be taken into account in studying the corresponding capillary phenomena.

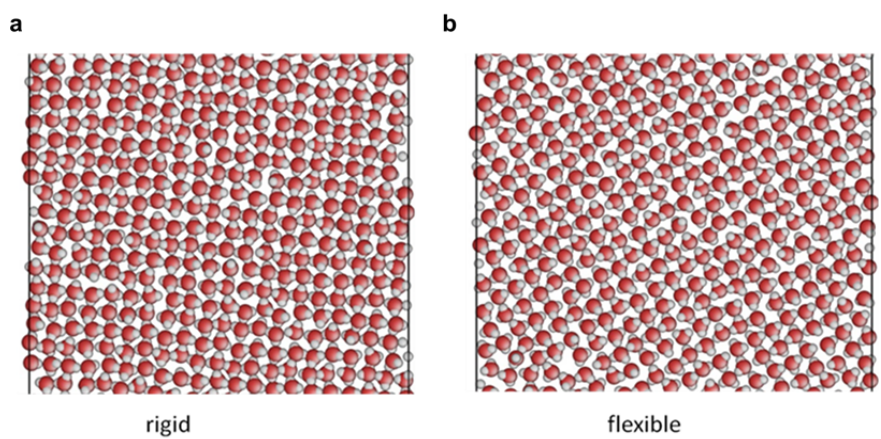

*Fig. S6. MD simulations of a monolayer of water confined in a 6.5 Å graphene capillary. Only one layer of water molecules can fit in. Red and grey circles represent oxygen and hydrogen atoms, respectively. Square ice is formed in this case, independent of whether the confinement is provided by rigid (a) or flexible (b) graphene sheets.*

#7. MD analysis of monolayer ice

For the narrowest (6.5 Å) graphene channel, a monolayer of water was found to form an ordered square lattice with $a$ = 2.80±0.01 Å for the rigid graphene confinement and 2.81±0.01 Å for the flexible one (Fig. S6), which agree well with the experimentally found $a$ = 2.83±0.03 Å. Within our MD accuracy, $a$ was found to be independent of pressure for all experimentally relevant $P$ from zero to 2 GPa. At higher pressures ($P$ >3 GPa), a puckered square ice started forming.

Simulations for the realistic confinement with freely-moving graphene sheets (Fig. S8d) yield the same square ice with the same lattice constant. To check if the SPC/E model provides sufficient accuracy, we have also used TIP4P/2005 model of water and again obtained the same square ice and the same $a$. Figs. S8d-f show small ice crystals that appear using the above two models with freely moving graphene sheets, which proves the robustness of our MD results with respect to different models.

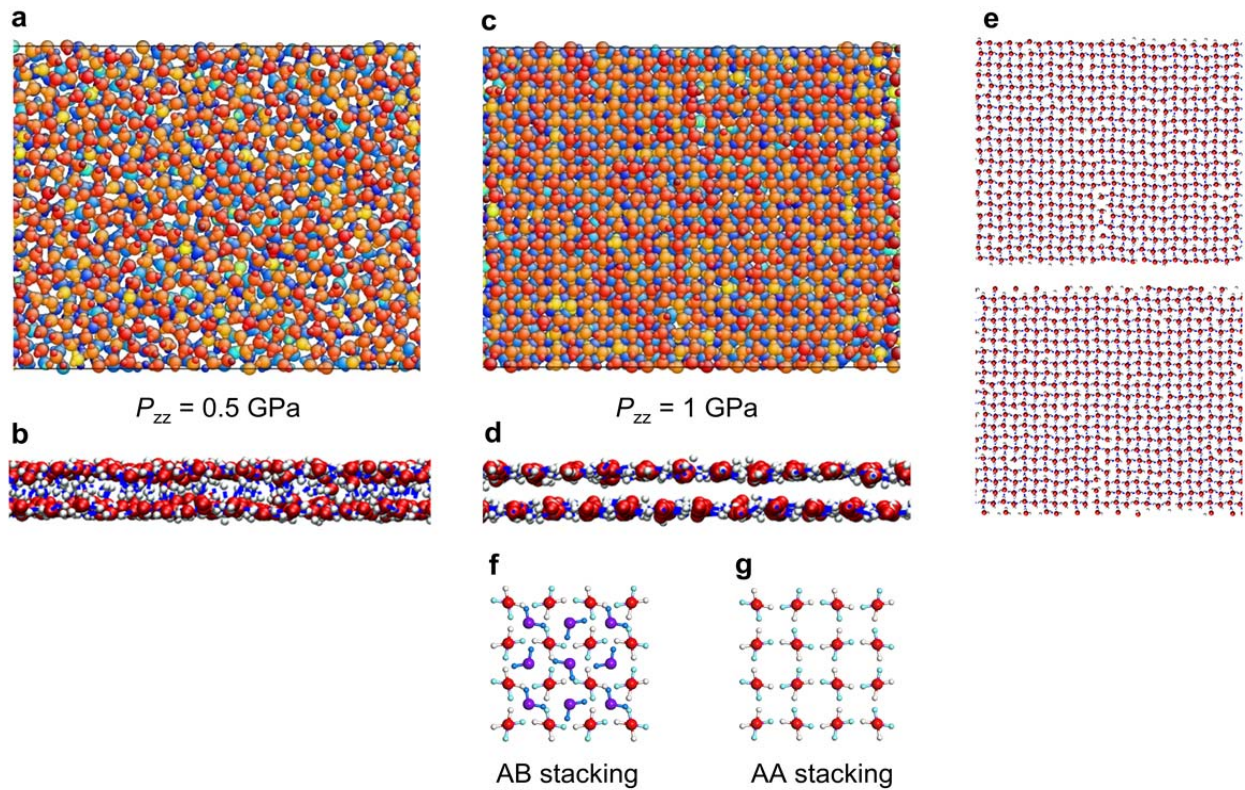

*Fig.S7. MD simulations of bilayer ice in a 9 Å high graphene capillary. (a,b)* Top (a) and side (b) views of a bilayer of water molecules formed at P =0.5 GPa. Different colors in (a) correspond to different vertical positions of water molecules in different layers: Dark blue and red circles mark the bottom and top positions, respectively; lighter colors correspond to intermediate positions. Red and white circles in (b) indicate oxygen and hydrogen atoms, respectively; short blue lines show hydrogen bonds. Although the water molecules are clearly arranged in two layers, no ordering is present and hydrogen bonds preserve their tetrahedral coordination as in bulk ices. *(c,d,e)* Same as in (a,b) but at P =1.0 GPa. Ordered bilayer ice is formed, with identical square lattices in the two layers, as illustrated by the snapshots of the top and bottom layers in (e). The ordering is accompanied by switching hydrogen bonds to in-plane coordination [see the short blue lines in (d)]. *(f,g)* Schematic illustration of what AB and AA stacking look like. The stacking found in (c) is AB.

#8. Simulated few-layer ice.

For the wider graphene capillaries (9 and 11.5Å) that accommodated two and three layers of water, respectively, no lattice formation was found at pressures below 1 GPa. A typical snapshot of the water molecule configuration in a bilayer at 0.5 GPa is given in Figs. S7a,b. In this case, water within each layer formed an amorphous structure, not dissimilar to bulk water or amorphous ice. Locally, water preserved the standard tetrahedral coordination with hydrogen bonds connecting the molecules both within and between the layers. As $P$ increased above $\approx$1 GPa, the simulations revealed an order-disorder transition accompanied by a sharp transformation in the hydrogen bonding between the two layers. One can see in Figs. S7b,d that hydrogen bonds switched to the in-plane configuration, effectively decoupling the two monolayers. At the same time, within each layer water molecules formed identical square lattices with the same lattice parameter as for the monolayer, $a$ =2.81±0.01 Å (Fig. S7e). A similar transition was found for the trilayer water ($h$ =11.5 Å) but at somewhat higher pressures, $P \approx$1.3 GPa (Fig. S9). The found values of $P$ at which the order-disorder transition occurs qualitatively agree with the estimated internal pressure induced by adhesion between graphene sheets.

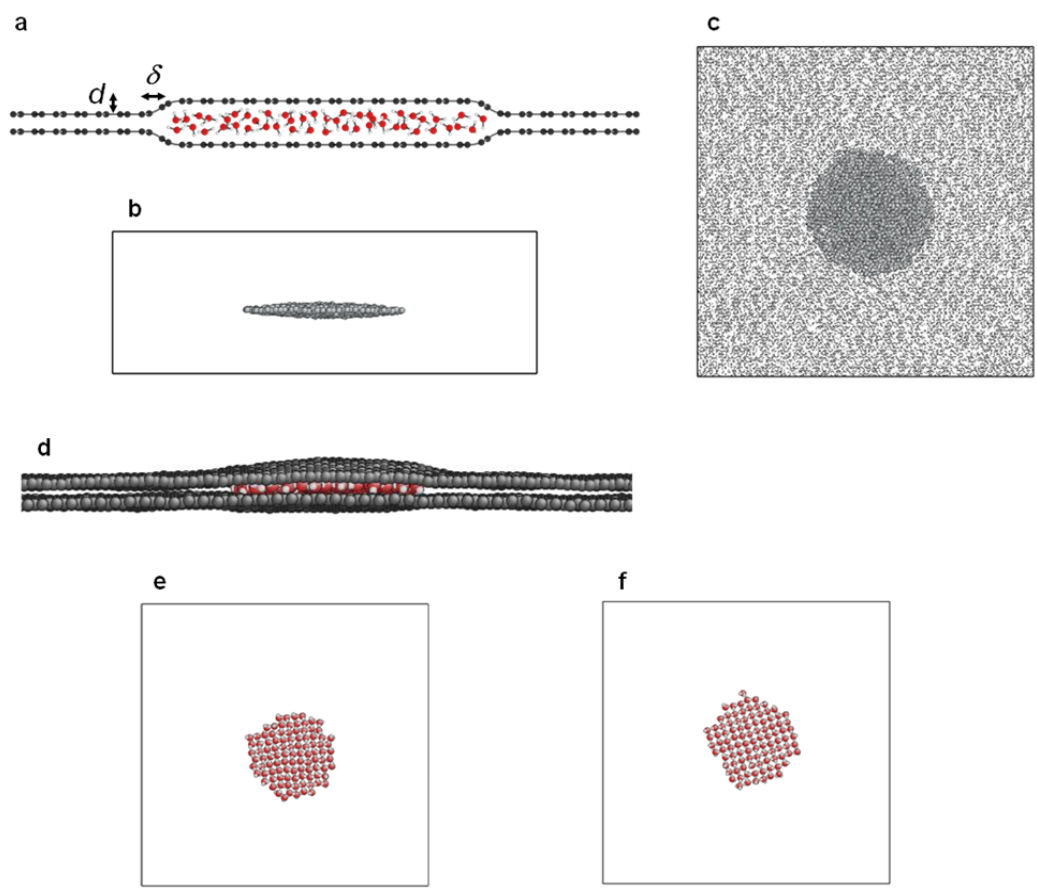

*Fig. S8. Van der Waals pressure acting on water confined between graphene sheets. (a) Schematic illustration of water confined by graphene. Due to adhesion between the graphene layers, water is squeezed into a 2D puddle. The shown parameters d and $\delta$ are used to estimate the vdW pressure (see text). (b,c) MD simulations of vdW pressure exerted on helium gas confined between two graphene sheets. In this simulation, a 'balloon' with 1,000 He atoms was encapsulated within freely-moving graphene sheets. (b) Side view of the helium balloon squeezed into a discus-shaped bubble at equilibrium. Graphene layers are not shown for clarity. (c) Top view of the He gas (dark grey) confined between graphene sheets (light grey). (d,e,f) MD simulations of 2D ice formed by a water nanodroplet (100 molecules) confined between two freely moving graphene sheets. Oxygen atoms are shown in red, hydrogen in light grey and graphene in dark grey. (d) Side and (e,f) top view at equilibrium. Simulations in (d,e) were performed using SPC/E model and in (f) using TIP4P/2005 model. For clarity, graphene is not shown in (e) and (f). The lattice parameter for the square ice in (e) and (f) is $\approx$2.8 Å within the modeling accuracy.*

The order-disorder transition was further confirmed by calculations of the per-atom potential energy for water molecules in the bi- and tri- layer ice, as shown in Figs. S9c,d. The sharp drop in the potential energy at $P \approx$ 1 and 1.3 GPa for the bilayer and trilayer, respectively, indicates ordering of the hydrogen bonds, which is similar to the known signature of the freezing transition for 3D water [48].

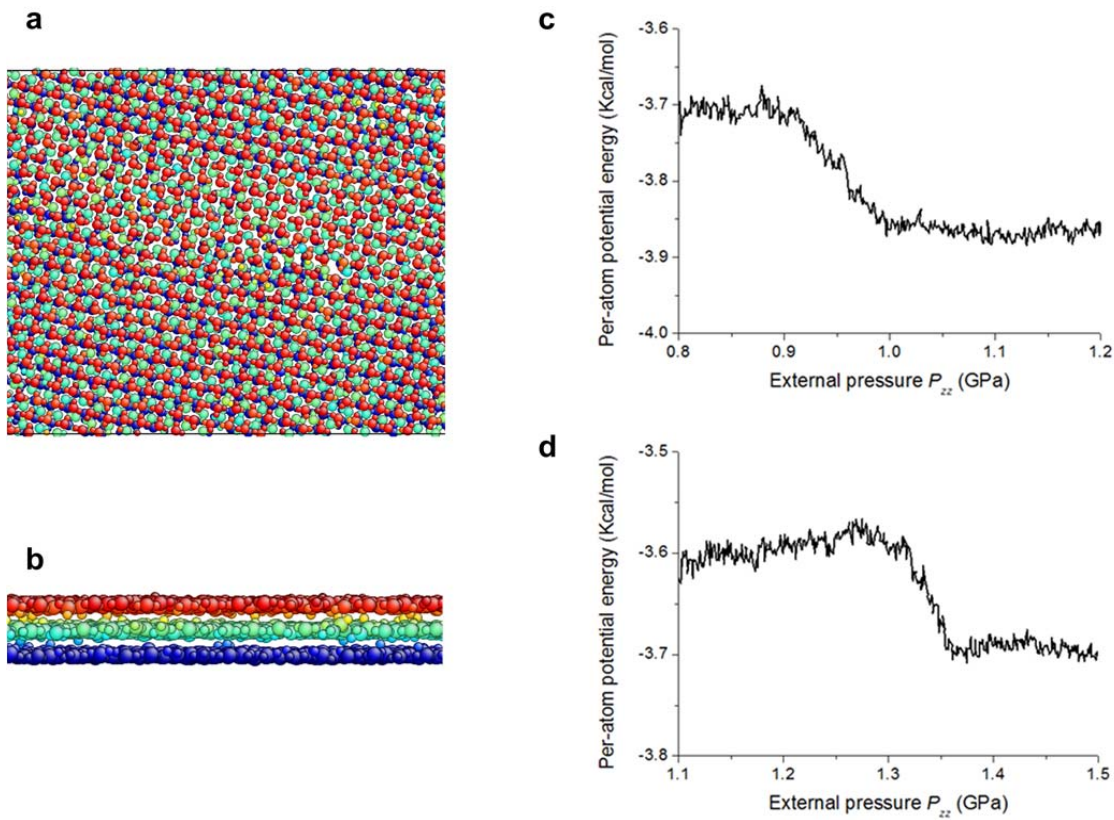

***Fig. S9. MD simulation of order-disorder transition. (a,b)*** *Top (a) and side (b) views of trilayer ice formed at P ≈1.4 GPa in the graphene capillary with h =11.5 Å . Different colors correspond to water molecules in different layers (see color coding in Fig. S7). No clear stacking sequence can be determined for this structure.* ***(c,d)*** *Potential energy per atom of 2D ice as a function of applied pressure for bilayer (c) and trilayer (d) cases.*

#9. Square ice inside non-graphene nanocapillaries.

To show that planar square ice could be common inside hydrophobic nanochannels other than those lined with graphene, we performed MD simulations of water confined between two generic walls. The latter were simulated using the Lennard-Jones potential in the form $E = 4\varepsilon[(\sigma/r)^{12} - (\sigma/r)^6]$ where $r$ is the perpendicular distance from the wall to water molecules. Parameters ε and σ are related, respectively, to the depth of the potential well and the distance at which the potential is zero. 1,000 water molecules were placed between such walls separated by 6.5 Å. Periodic boundary conditions were applied in the directions parallel to the walls, and the simulations were carried out using an isothermal-isobaric ensemble as described above. To model capillaries with different hydrophobicity, ε was varied from 0.01 to 0.2 kcal/mol. This covers a wide range of surfaces from super-hydrophobic to weakly hydrophilic. Figs. S10a-d illustrate the contact angles for such surfaces. Square ice formed inside all the modelled nanochannels at room temperature, independently of their ε, and its lattice parameter is found to be the same within ±1% (Figs. S10e-h). Furthermore, we checked that a crystallographic structure of hydrophobic walls was not important for the occurrence of square ice. To this end, a monolayer of water was confined between artificial carbon walls with lattices other than the hexagonal one of graphene (for example, square lattice). Again, we observed the same square ice with the same *a*. This allows us to conclude that the square ice is likely to be common inside hydrophobic channels that can accommodate only a few monolayers of water.

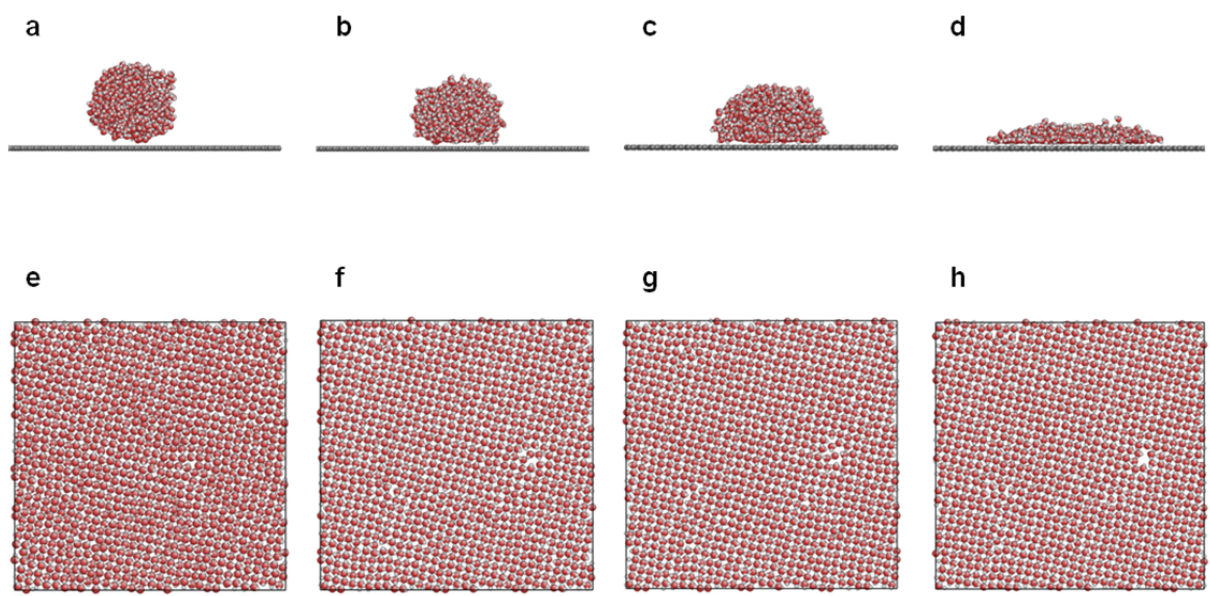

*Fig. S10. MD simulations of monolayer ice confined between different hydrophobic surfaces. (a-d)* *Water droplets on surfaces with different water-surface interactions, $\varepsilon$ = 0.01, 0.05, 0.1 and 0.2 kcal/mol, respectively. The changes in the relative strength of water-water and water-surface interactions are reflected in different contact angles. For comparison, $\varepsilon$ for graphite-water interaction is 0.07 kcal/mol.* ***(e-h)*** *Monolayer ice confined between surfaces with values of $\varepsilon$ corresponding to (a-d), respectively. In this case, the modeled walls are generic and do not have a discrete atomic structure, i.e., their potential is uniform. h =6.5 Å, same as for monolayer ice in the graphene channel in Fig. S6. Square ice is found to form in all the cases and has the same a.*

Further work is necessary to understand how different modelling parameters and geometries may influence the outcome of MD simulations and, for example, to explain the AA stacking for bi-/tri-layer square ice and consequences of the terraced graphene confinement observed experimentally.